\documentclass[12pt,a4paper,dvips]{article}

\usepackage{graphicx}
\usepackage{amsbsy}
\usepackage{amssymb}

\begin{document}

\newcommand{\be}{\begin{equation}}
\newcommand{\ee}{\end{equation}}
\newcommand{\bea}{\begin{eqnarray}}
\newcommand{\eea}{\end{eqnarray}}
\newcommand{\tr}{\,\hbox{tr }}
\newcommand{\Tr}{\,\hbox{Tr }}
\newcommand{\Det}{\,\hbox{Det }}
\newcommand{\fslash}{\hskip-.2cm /}
\begin{center}
{\bf \Large Black Holes Without Singularities}
\end{center}
\vspace*{0.2cm}
\begin{center}
A. Bogojevi\'c
\end{center}
\begin{center}
{\it Institute of Physics\\
P.O.B. 57, Belgrade 11001, Yugoslavia}
\end{center}
\begin{center}
D. Stojkovi\'c
\end{center}
\begin{center}
{\it Department of Physics\\
Case Western Reserve University, Cleveland, OH 44106}
\end{center}
\vspace*{0.1cm}
\begin{abstract}
{\small We study the properties of a completely integrable deformation of the 
CGHS dilaton gravity model in two dimensions. The solution is shown to
represent a singularity free black hole that at large distances
asymptoticaly joins to the CGHS solution.}
\end{abstract}
\vspace*{0.1cm}

One of the fundamental unsolved problems in theoretical physics is the
unification of quantum theory and gravity. One of the reason why this is
so difficult stems from the complicated nonlinear structure of the
equations of general relativity. These equations are much simpler
in lower dimensions. For this reason there has recently been much activity
related to the quantization of gravity in two and three dimensions. One of
the most important results in 2d was the exactly solvable dilaton gravity
model constructed by Callan, Giddings, Harvey and Strominger \cite{cghs}.
The CGHS model has 2d black hole solutions that are remarkably similar to
the Schwarzschild solution of general relativity. Of the four fundamental
interactions in nature, gravity is by far the weakest. For this reason,
we can hope to see quantum effects only in the vicinity of classical
singularities. Penrose and Hawking have shown that these singularities
are endemic in general relativity. The general belief is that quantization
will rid gravitation of singularities, just as in atomic physics it got
rid of the singularity of the Coulomb potential. If this is indeed the
case, then there must exist a non-singular gravitational effective action
whose classical equations encode the full quantum theory. This effective
action must have the Planck length $L_\mathrm{Planck}$ in it as an input
parameter. For $L\gg L_\mathrm{Planck}$ the effective model must be
indistinguishable from the classical gravity action. The search for such
an effective model parallels Landau's treatment of phase transitions in
ferromagnets. Landau chose (the simplest) effective action (Gibbs potential
in statistical mechanics parlance) that led to a qualitatively correct
discription of phase transitions.

An important result in dilaton gravity has been the work of Louis-Martinez
and Kunstatter \cite{lmk}, who reduced the solution of the general dilaton
gravity model to the solution of two ordinary integrals, i.e. to two
quadratures. In a previous paper \cite{bs} we used their proceedure to
construct a dilaton model that yields a black hole without a singularity.
In this paper we will review the central results of this derivation,
construct the deformed CGHS model and show that it leads to a maximal
curvature proportional to $L_\mathrm{Planck}^{-1}$.

The action of all dilaton gravity models can be put into the general form
\be
\label{general action}
S=\int d^2x \sqrt{-g}\left[\,{1\over 2}\,g^{\alpha\beta} 
\partial_\alpha \phi 
\partial_\beta \phi - V(\phi )+D(\phi)R\right]\ .
\ee
The potentials $V(\phi)$ and $D(\phi)$ classify all the possible models. 
Performing a conformal scaling of the metric
$\tilde g_{\alpha\beta}=e^{-2F(\phi)}g_{\alpha\beta}$, where the scaling 
factor $F(\phi)$ satisfies $F'=-1/4(D')^{-1}$ we can put the action into
the simplified form
\be 
\label{simplified action}
S=\int d^2 x\sqrt {-\tilde g}
\left[ \tilde\phi \tilde R-\tilde V (\tilde \phi)\right]\ ,
\ee
where $\tilde R$ is the scalar curvature corresponding to 
$\tilde g_{\alpha\beta}$, and we have introduced the new dilaton field and
potential according to $\tilde \phi = D(\phi)$ and 
$\tilde V(\tilde\phi)=e^{2F(\phi)}V(\phi)$.
This form of the dilaton gravity action is obviously much easier to work
with since we have lost the kinetic term for the dilaton field.

A well known property of two dimensional manifolds allows us to localy,
i.e. patch by patch, choose conformally flat coordinates for which
$\tilde g_{\alpha\beta}=e^{2\rho}\,\eta _{\alpha \beta}$. 
Louis-Martinez and Kunstatter \cite{lmk} have shown that we can choose a
coordinate system in which the solution of the general dilaton model
is static and given by
\bea
\label{second quadrature} 
x&=&-2\int {d\tilde\phi\over W(\tilde\phi)+C}\\
e^{2\rho}&=&-\,{C+W(\tilde\phi)\over 4}\ ,
\eea
where the pre-potential $W(\tilde\phi)$ is given by 
${dW\over d\tilde\phi}=\tilde V(\tilde\phi)$, and $C$ is an invariant.
It is easy to show that $C<0$, and that without loss of generality we can
choose $C=-1$. As we can see, the above solution is given in terms of two
quadratures: the first connecting $F$ and $D$, and second one given in
(\ref{second quadrature}). A given model is completely integrable only if
we can calculate both quadratures in closed form.

The CGHS model is an example of a completely integrable dilaton gravity
model. The standard form of the CGHS action is
\be
S=\int d^2x\sqrt{-g}\,e^{-2\varphi}\,(R+4g^{\alpha\beta}
\partial_\alpha\varphi\partial_\beta\varphi +4\lambda^2)\ .
\ee
The simple field redefinition $\phi=\sqrt{8}\, e^{-\varphi}$ puts this into
the general form for dilaton gravity actions given in (\ref{general action}).
Further, a conformal scaling with $F(\phi)=-\ln\phi$ gives us the simplified
form of the CGHS action
\be
S=\int d^2x\sqrt{-\tilde g}
\left(\tilde\phi\tilde R+{1\over 2}\,\lambda^2\right)\ .
\ee
The CGHS model is completely integrable. A simple application of the 
Louis-Martinez and Kunstatter proceedure gives us the general solution. For
the scalar curvature we find $R=-32\,A^{-1}$, where we have introduced
$A=8/\lambda^2\,\left(e^{{\lambda^2\over 4}x}-1\right)$. The metric for the 
general dilaton model, given in terms of $F$ and $\rho$, is simply
$ds^2=e^{2(F+\rho)}(-dt^2+dx^2)$. In the case of CGHS we get
\be
e^{2(F+\rho)}={\lambda^2\over 64}\,
{e^{{\lambda^2\over 4}x}\over e^{{\lambda^2\over 4}x}-1}\ ,
\ee
which vanishes for $x=-\infty$. For stationary metrics the equation
$g_{00}=0$ determines the horizon. Therefore, in these coordinates the CGHS
black hole has a horizon at $x=-\infty$. The curvature, on the other hand,
is well behaved at this point. As with the Schwartzschild black hole one can
now find coordinates which are well behaved at the horizon. In this way one
finally obtains information about the global character of the manifold.

We now proceed to construct a new dilaton gravity model that satisfies the
following requirements: it is completely integrable, for $x\to\infty$ it
goes over into the CGHS model and is singularity free. As we have seen,
dilaton gravity models are specified by giving the two potentials $D(\phi)$
and $V(\phi)$. It is very difficult to see how one should deform these
potentials from their CGHS form in order to satisfy the above criteria.
Note, however, that the models are also uniquely determined by giving
$F(\phi)$ and $\tilde V(\tilde\phi)$. This is much better for us since
we have now untangled the two integrability requirements: $F(\phi)$
determines the first quadrature and $\tilde V(\tilde\phi)$ the second.
Deformations of a given model correspond to changes of both of these
functions. In this paper we will look at a simpler problem. We shall keep
$\tilde V(\tilde\phi)$ fixed, i.e. it will have the same value as in
the CGHS model. We will only deform $F(\phi)$. By doing this we are
guaranteed that the second (and more difficult) quadrature is automatically
solved. From our second requirement we see that for large $x$ the dilaton
field $\phi(x)$ must be near to its CGHS form. Specifically, $x\to\infty$
corresponds to $\phi\to\infty$. Thus, our second requirement imposes that
for $\phi\to\infty$ we have $F(\phi)\to-\ln\phi$.
$F(\phi)$ must also be such that the first quadrature is exactly solvable.
To do this we choose
\be
\label{F deformed}
F(\phi)=-{1\over\alpha}
\ln\left({1+\beta \phi^\alpha\over\beta}\right)\ ,
\ee
with $\alpha>0$. The $\alpha$ and $\beta$ values parametrize our class of
deformations. The first quadrature gives
\be
\label{D deform}
D(\phi)=\left\{ 
\begin{array}{ll}
{1\over 8}\phi^2+{1\over 4\beta}\ln\phi&\hbox{  for }\alpha=2\\
{1\over 8}\phi^2+{1\over 4\beta(2-\alpha)}\phi^{2-\alpha}&
\hbox{  for }\alpha\neq 2\ .
\end{array}
\right.
\ee
On the other hand, the potential $V(\phi)$ is now simply
\be
V(\phi)=-{1\over 2}\lambda^2
\left({1+\beta\phi^\alpha\over\beta}\right)^{2\over\alpha}\ . 
\ee
The choice of $\alpha$ corresponds to a choice of explicit model, while
$\beta$ just sets a scale for the dilaton field. Rather than work here with
the general deformed model we will now concentrate on the simplest model in
this class; the one corresponding to the choice $\alpha=4$. The action for
this model is
\be
\label{deformed action}
S=\int d^2x\sqrt{-g}
\left({1\over 2}\,g^{\alpha\beta}\partial_\alpha\phi\partial_\beta\phi
+{1\over 2}\,\lambda^2
\left({1+\beta\phi^4\over\beta}\right)^{1\over 2}
+{1\over 8}\left(\phi^2-{1\over\beta\phi^2}\right)R\right)\ .
\ee
Note that for $\beta\to\infty$ this goes over into the action of the CGHS
model. As we have seen, $\beta$ is just a scale for $\phi$, hence, this is
just a re-statement of our second requirement. From our construction we see
that (\ref{deformed action}) corresponds, for each finite value of $\beta$,
to a model that satisfies our first two requirements. All that is left is
to check that the theory is indeed free of singularities. Being in two
dimensions all that we need to check is the scalar curvature. A simple
but tedious calculation now gives
\bea
\label{deformed curvature}
R&=&\sqrt{2}\,\lambda^2\left({1\over\beta}+A^2\right)^{-{7\over 4}}
\left(A+\sqrt{{1\over\beta}+A^2}\,\right)^{1\over 2}\cdot\nonumber \\
& &\qquad\cdot\left\{{16\over\beta\lambda^2}+{3\over\beta}A-
{8\over\lambda^2}A^2+
\left({1\over\beta}-{8\over\lambda^2}A\right)
\sqrt{{1\over\beta}+A^2}\,\right\}\ .
\eea
For $\beta\to\infty$ we indeed find that $R$ goes over into the CGHS
result. From (\ref{deformed curvature}) we see that the curvature of the
deformed CGHS model is indeed not singular.
\begin{figure}[!ht]
  \centering
  \includegraphics[width=7cm]{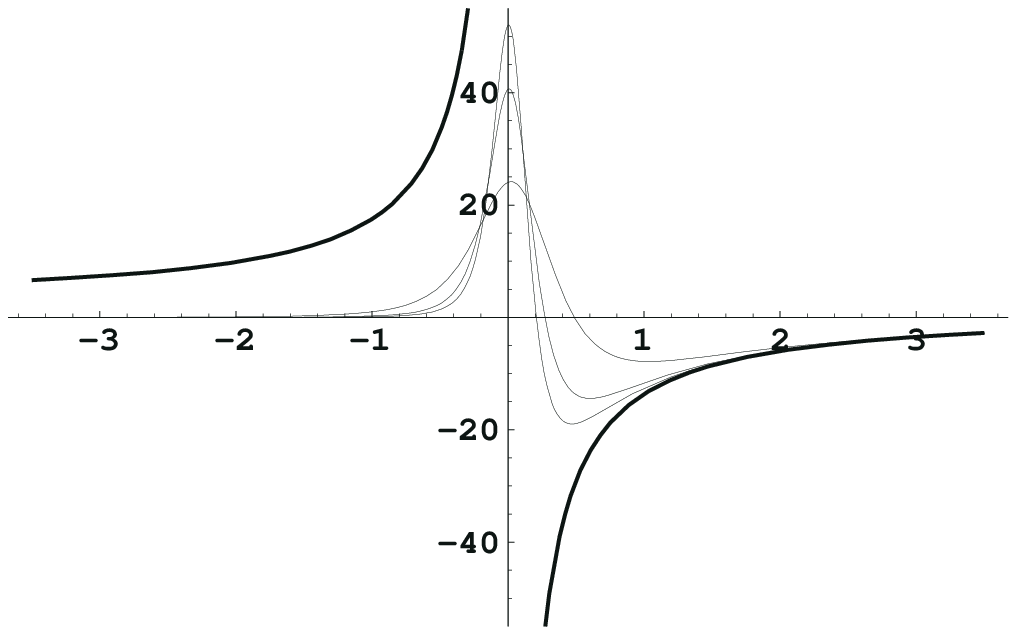}
  \caption{
$R(x)$ for the CGHS model (thick line) and deformations with $\beta=1$,
3 and 5. As $\beta$ increases the deformations for $x>0$ join CGHS. The
plot is
for $\lambda^2=1$}
\end{figure}
As may be seen in Figure~1, the deformed model has maximal curvature at
$x=0$. Its value is
\be
\label{R max}
R_\mathrm{max}=\sqrt{2}\,\left(16\beta^{1\over 2}+\lambda^2\right)\ .
\ee
At right infinity the deformed model tends to the CGHS result. On the other
hand, at left infinity both the CGHS model and its deformation tend to a
de Sitter space $R=\Lambda$. However, for CGHS we have $\Lambda=4\lambda^2$,
while for the deformed model the constant is a complicated function of
$\beta$ and $\lambda$. Rather than writing it out let us only give the
result for large $\beta$ when we have
$\Lambda=2^{-10}\lambda^8\beta^{-{3\over 2}}$. 
We have just determined that the $x\to -\infty$ and $\beta\to\infty$ limits
do not commute. Therefore, imposing that our model joins to CGHS at right
infinity doesn't automaticaly guarantee a similar joininig at left infinity.

We are now in the position of trying to interpret the meaning of our
deformed CGHS model. Obviously, one possibility is to think of
(\ref{deformed action}) as the classical action
of a model with scale $1\over\beta$. However, it seems more natural to
interpret our model as an effective action. $1\over\beta$ then naturaly
comes about from quantization, while $\beta\to\infty$ corresponds to the
semi classical limit. Our model should thus be the effective action
corresponding to the quantization of the CGHS model. Quantization gives
$S\sim\hbar$, and essentialy dimensional analysis (in units $G=c=1$)
gives $\phi^2\sim\hbar$, as well as ${1\over\beta}\sim\hbar^2$.
Therefore, if we are to interpret our model as an effective action then
$\beta=\kappa\hbar^{-2}$, where $\kappa$ is a constant of the order of
unity. We see then that the maximal curvature (\ref{R max}) is proportional
to ${1\over\hbar}$, i.e. represents a non-perturbative effect.
Expanding our model in $\hbar$ we find
\be
S_\mathrm{eff}=S_\mathrm{cghs}-{1\over 8\kappa}\,\hbar^2\,
\int d^2x\,\sqrt{-g}\,\left(R-2\lambda^2\right)\phi^{-2}+o(\hbar^4)\ .
\label{eff}
\ee
The leading correction to CGHS is of the form of the Jackiw-Teitelboim
action for 2d gravity. It would be very interesting to get this result
by quantizing some fundamental 2d theory. To do this we would need to
start from the CGHS model coupled to some matter fields. We would then
have to integrate out the matter. The last step would be to calculate
the effective action. It is probably impossible to do this exactly,
however, we could hope to do this perturbatively and compare with
(\ref{eff}).

\end{document}